\documentclass[twocolumn]{revtex4-1}

\usepackage{graphicx}
\usepackage{amsfonts}
\usepackage{amssymb}

\newcommand{\ts}{\tilde{\sigma}}
\newcommand{\hT}{\hat{T}}
\newcommand{\s}{\sigma}
\newcommand{\be}{\begin{equation}}
\newcommand{\ee}{\end{equation}}
\newcommand{\bea}{\begin{eqnarray}}
\newcommand{\eea}{\end{eqnarray}}

\begin{document}
\title{A Finite State Model for Time Travel}

\author{Hwee Kuan Lee}

\affiliation{Bioinformatics Institute, 30 Biopolis Street, \#07-01, Matrix, Singapore 138671}

\begin{abstract} 
A time machine that sends information back to the past may, in principle, be built
using closed time-like curves. However, the realization of a time machine must be 
congruent with apparent paradoxes that arise from traveling back in time. Using a
simple model to analyze the consequences of time travel, we show that
several paradoxes, including the grandfather paradox and Deutsch's unproven theorem
paradox, are precluded by basic axioms of probability. However, our model does
not prohibit traveling back in time to affect past events in a self-consistent manner.
\end{abstract}

\maketitle

\section{Background}

The possibility of building a time machine has been proposed by many 
authors
~\cite{friedman,gott,godel,bonnor,morris,politzer,boulware,hartie,politzer2,deutsch,
novikov,lloyd,pegg,svetlichny}.
Two common approaches are through closed time-like curves 
(CTC)~\cite{friedman,gott,godel,bonnor,morris,politzer,boulware,hartie,politzer2} and
quantum phenomena~\cite{deutsch,lloyd,pegg,svetlichny}. Although
the general theory of relativity allows for CTCs, it is not clear if the laws of
physics permit their existence~\cite{hawking,carroll,deser,carroll2}. Hence the 
possibility of traveling back to the distant past remains an open question.
Paradoxical thought experiments have been devised to suggest that traveling back in
time may lead to violations of causality, and hence is not possible. The most famous
paradox is the grandfather paradox, in which an agent travels back in time to kill
his grandfather before his father was conceived. In this case, the agent will not
exist at the current time and hence cannot travel back in time to kill his
grandfather. An alternative version of the grandfather paradox is autoinfanticide, where
an agent travels back in time to kill himself as an infant. This paradox plays a
central role in the argument against traveling back in time. 
Another paradox is the Deutsch's unproven theorem paradox~\cite{lloyd}, in which an
agent travels back in time to reveal the proof of a mathematical theorem. The proof is
then recorded in a document that the agent reads in future time. 
Another version of Deutsch's unproven paradox is
what we call the chicken-and-egg paradox. A hen travels back in time to lay an
egg. The egg hatches into the hen herself. Without the egg, the hen would not exist but
without the hen traveling back in time, the egg would not be laid. 

In this paper, a simple model is used in an attempt to solve time travel
paradoxes and help set the logical foundations of traveling back in time. 
Our approach is quite different from approaches that focus on how a time machine can
be built (in  principle)~\cite{lloyd}. We suppose that a time machine can be
built and then analyze what could be possible (or impossible) in time travel. We use a
simple directed cyclic graph to explain causal relationships in different scenarios of
time travel. 
Our conclusion is that, assuming traveling back in time is feasible, an agent
who travels back in time is unable to kill himself although he may be able to alter
the past in other ways; in a self-consistent manner.

The self-consistency principle was proposed by Wheeler and Feynman~\cite{feynman},
Novikov {\it et al}~\cite{novikov} and Lloyd {\it et al}~\cite{lloyd}. 
It states that traveling back in time may be possible, but it cannot happen in
a way that violates causality. Causality in this case includes events that happen in
the future affecting the past. This principle precludes time travel paradoxes but does
not forbid traveling back in time. Due to space limitations, the reader is referred 
to ~\cite{feynman,novikov,lloyd} for detailed discussion of the self-consistency
principle.

\section{Model}

Our model can be considered as a simple case of graphical models.
Graphical models have been extensively studied and are applicable in many fields
such as in econometric models, social sciences, artificial intelligence and even in
medical studies. Publications on graphical models are so numerous that we can only
provide a non-exhaustive list
~\cite{richardson1997,richardson1996,schmidt,spirtes,lacerda,pearlbk,rebane1987,lauritzen,
  lauritzen1,salmon,morgan,spirtesbk,cooper1991}.
Although directed acyclic graphs have been at the center stage of graphical models,
directed cyclic graphical models have also received significant attention
~\cite{schmidt,richardson1997,richardson1996,spirtes,lacerda}. 
Two important components in graphical models are intervention
and the {\it do calculus}. The theory of
graphical models has few constraints built in on what is physically possible. This
leaves the theory very general.

\begin{figure}[h!]
\begin{picture}(200,0)(0,0)
\put(0,0){$
\s_1 \rightarrow \s_2 \rightarrow \s_3 
\rightarrow \s_4
\cdots
\rightarrow \s_i
\cdots
\rightarrow \s_k
\cdots
\rightarrow \s_n 
$}
\end{picture}
\caption{A simple graphical model for a Markov Chain}
\label{fig:mc}
\end{figure}

We use a simple directed cyclic graph to study traveling back in time.
First, we build constraints into our model as follows. Consider
physical states evolving on a timeline as shown in Fig. \ref{fig:mc}. The graph is a
one dimensional chain, and branching is excluded. Traveling back in time introduces a
loop as in Fig. \ref{fig:loop}. We do not include intervention and {\it do
calculus} because this enables us to simplify
our analysis, while capturing the important physics for a closed system.

\begin{figure}[h!]
\begin{picture}(200,20)(0,0)
\put(0,5){$
\s_1 \rightarrow \s_2 \rightarrow \s_3 
\rightarrow \s_4
\cdots
\rightarrow \s_i
\cdots
\rightarrow \s_k
\cdots
\rightarrow \s_n $}
\put(112,-10){\vector(0,1){12}}
\put(112,-10){\line(1,0){33}}
\put(145,-10){\line(0,1){12}}
\end{picture}
\caption{A simple cyclic graph to model traveling back in time from $t=k$ to $t=i$.}
\label{fig:loop}
\end{figure}

At each time $t$, the state of the system $\s_t$ is a random variable. 
Time is also discretized and the arrows connect events at neighboring times 
$\s_t \rightarrow \s_{t+1}$. The probability of transition from $\s_t$ to $\s_{t+1}$
is given by $T_{t+1}(\s_{t+1}|\s_t)$. In this case, the conditional probabilities can
be interpreted as a transition matrix, and the graph as a Markov
Chain. The following assumptions are used based on physical considerations:
\begin{enumerate}
\item The statistical time flows in the same direction as the physical time.
\item Local normalization constraint is enforced, i.e. 
$\sum_{\s_{t+1}} T_{t+1}(\s_{t+1}|\s_t)=1$. Given that the system is in a state
$\s_t$ at time $t$, the system has to take on a state at $t+1$. In general, we can
condition on more than one variable, e.g. $T_{t+1}(\s_{t+1}|\s_i,\s_j,\cdots)$,
then the local normalization condition is 
$\sum_{\s_{t+1}} T_{t+1}(\s_{t+1}|\s_i,\s_j,\cdots)=1$.
\item Basic probability axioms are satisfied. Let $A_i$ be a set of states and 
$P(A_i)$ be its probability measure, then,
\be
0 \leq P(A_i) \leq 1
\label{eq:prange}
\ee
\be
P(\Omega) = 1, \mbox{\hspace{0.5cm}}
\label{eq:pnorm}
\ee 
\be
P(A_i \cup A_j) = P(A_i) + P(A_j), 
\ee
$\Omega$ is the set of all possible states and $A_i$ and $A_j$ are mutually
exclusive. Clearly, for discrete events if $\s_i\in \Omega$ and $\s_j \in \Omega$, 
$\s_i \neq \s_j$, then $P(\s_i \cup \s_j) = P( \s_i) + P(\s_j)$. Here, we use a
shorthand notation $\s_i \equiv \{ \s_i\}$.
\end{enumerate}

A sequence of states $\pi_n$ is shown in Fig. \ref{fig:mc}. If the set of all
possible states is given by $\Omega$, then the set of all possible sequences is given
by $\mathbf \Pi = \Omega^n$. The probability of obtaining $\pi_n$ is,
\be
P_{mc}(\pi_n) = p(\s_1) T_2(\s_2|\s_1) T_3(\s_3|\s_2) \cdots 
T_n(\s_n|\s_{n-1}) 
\label{eq:mcseq}
\ee
$p(\s_1)$ is the probability of sampling the initial state $\s_1$. The conditional
probabilities encode the physics of how the system evolve from state to state.
It can be shown that for $P_{mc}(\pi_n)$, basic axioms of probabilities hold.

In the case of traveling back in time, the causal relationship has an arrow
that loops back into the past (Fig. \ref{fig:loop}).
To model traveling back in time, we condition on two states instead of one, 
$\hT_i(\s_i | \s_{i-1}, \s_k)$ where $\s_k$ is an event in the future with
respect to time $i$. In this case, 
\be
P(\pi_n) = p(\s_1)T_2(\s_2|\s_1)\cdots \hT_i(\s_i|\s_{i-1},\s_k)
\cdots T_n(\s_n|\s_{n-1})
\label{eq:mcloop}
\ee
All the conditional probabilities $T_j(\s_j|\s_{j-1})$ are the same as in 
Eq. (\ref{eq:mcseq}) except for $\hT_i(\s_i|\s_{i-1},\s_k)$. Making such a
generalization is non-trivial because we need to check that the basic axioms of
probabilities continue to hold.
At this point, we would like to emphasize some key points that are important in this paper,
\begin{enumerate}
\item Time travel consists of sending a signal back to the past.
The signal causes an effect only at one
time point $t=i$ as in Fig. \ref{fig:loop}. The signal could contain a set of
instructions to carry out some tasks or be an agent that travels back in time.
\item The conditional probabilities $T_j$, $j=1,2,\cdots$, $j\neq i$ in 
Eq. (\ref{eq:mcseq}) are determined by the physics of how the system evolves forward in
time.
\item The term $\hT_i(\s_i|\s_{i-1},\s_k)$ is special as it is the only term in 
Eq. (\ref{eq:mcloop}) that encodes the effects of traveling back in time. 
\item
Our framework is probabilistic, in which many sequences of states can happen with
non-zero probability, in contrast to a deterministic view where only one sequence is
possible. Given any sequence $\pi_n$, its probability of occurrence can be calculated
using Eq. (\ref{eq:mcloop}). 
\item A paradox be represented by many different sequences
of states. Our objective is to show that either all these sequences happen with zero 
probability, or they result in violation of the basic axioms of probability.
\end{enumerate}

Consider $\hT_i$ to be a function of three discrete variables, 
$\s_{i-1},\s_i$ and $\s_k$. This function has to satisfy,
\be
0 \leq \hT_i(\s_i | \s_{i-1},\s_k)\leq 1
\label{eq:thatrange}
\ee
\be
\sum_{\{ \pi_n\} } P(\pi_n) = 1
\label{eq:pinorm}
\ee
\be
\sum_{\s_i} \hT_i(\s_i | \s_{i-1},\s_k) =  1
\label{eq:thatnorm}
\ee
The first two conditions are analogous to Eq. (\ref{eq:prange}) and (\ref{eq:pnorm}).
The last condition is the local normalization condition.
Eq. (\ref{eq:pinorm}) can be reduced to,
\be
\sum_{\s_i,\s_k} 
\hT_i(\s_i | \ts_{i-1},\s_k)
V(\s_k|\s_i ) = 1
\label{eq:TV}
\ee
$V(\s_k|\s_i)$ is the conditional probability of $\s_k$ given $\s_i$ summed over
all possible intermediate states $\s_{i+1}\cdots \s_{k-1}$. Detailed derivation of
Eq. (\ref{eq:TV}) is given in Appendix A.
This is an important equation. We will use this equation together with  
Eq. (\ref{eq:thatrange}) and (\ref{eq:thatnorm}) to show that the grandfather 
paradox, Deutsch's unproven theorem paradox and chicken-and-egg paradox have to
be precluded in time travel.

\subsection{Two-state system}

For a two-state system, $\s$ takes the values $\{0,1\}$.
Using Eq. (\ref{eq:TV}) and (\ref{eq:thatnorm}) and summing
over four combinations $\s_{i+1}, \s_k \in \{0,1\}$,  we obtain,
\be
[\hT_i(0|\ts_{i-1},1) - \hT_i(0|\ts_{i-1},0)] [ V(1|0) - V(1|1) ] = 0
\label{eq:2state}
\ee
We must have $V(1|0) = V(1|1)$ or $\hT_i(0|\ts_i,1) = \hT_i(0|\ts_i,0)$. For the case
when $V(1|0) \neq V(1|1)$, the transition matrix $\hT_i$ does not depend on
$\s_k$. In this case, the backward loop in Fig. \ref{fig:loop} has no effect. We
can't change the  probability distribution of the past. For the case 
$V(1|0) = V(1|1)$, we could have $\hT_i(0|\ts_{i-1},1) \neq \hT_i(0|\ts_{i-1},0)$ and
the transition probabilities at $t=i$ could be affected by a signal from future time
($t=k$).

\subsection{Grandfather paradox in a two-state system}

The grandfather paradox can be used to illustrate the physical implications of
Eq. (\ref{eq:2state}). The basic assumptions we will use are (i) resurrection is
impossible, and (ii) basic axioms of probabilities must be satisfied.

Consider an agent sending a signal back in time to kill himself.
Let us denote the dead state as $\s=0$ and alive state as $\s=1$. No resurrection 
implies that $V$ is of the form,
$
V = \left( \begin{array}{cc}
1 & \beta^*  \\
0 & \beta  \end{array} \right),
$
$\beta^*=1-\beta$. Let 
$\hT_i(\s_i|\ts_{i-1},1) = S(\s_i|\ts_{i-1})$ be the transition probabilities 
for the scenario in which the
agent sends a signal from the future to kill himself.
Let $\hT_i(\s_i|\ts_{i-1},0) = N(\s_i|\ts_{i-1})$ be the transition probabilities for 
the sequences of events
the agent is dead at $t=k$ and hence no signal is sent from the future to kill
himself. 
Hence $S$ (the ``killing" matrix) and $N$ are of the form,
\be
S = \left( \begin{array}{cc}
1 & 1 \\
0 & 0 \end{array} \right) \mbox{\hspace{.6cm}}
N = \left( \begin{array}{cc}
1 & b^* \\
0 & b \end{array} \right)
\ee
$b^* = 1-b$ is the probability of dying at $t=i$. Substituting values of $N$, $S$
and $V$ into Eq. (\ref{eq:2state}), we obtain
$ [1-b^*]\beta= 0$.
Either $b^*=1$ or $\beta=0$. When $b^*=1$ then $N=S$, the agent dies at $t=i$ with 
probability 1. If
$\beta=0$, the agent dies sometime between $t=i$ and $t=k$ with probability 1.
In either case, the scenario in which the agent is alive at $t=k$ and thus able to 
send the signal occurs with zero probability.
Note that we are analyzing probabilities rather than specific events.

The conclusion comes about because resurrection is impossible ($V(1|0)=0$). Suppose
resurrection is possible, $V(1|0)=\alpha^*<1$, the paradox goes away when 
$\alpha^*=\beta$. Intuitively, if we allow resurrection, the agent could send a 
signal back in time from $t=k$ to kill himself at $t=i<k$. Between the time $t=i$ 
and $t=k$, the agent is resurrected and hence could again send the signal at $t=k$. 
There is no contradiction in this case.

Another way to resolve the paradox is to relax the assumption that the agent always
succeeds to kill himself. In this case, the matrix $S$ is 
$\left( \begin{array}{cc}
1 & \lambda^* \\
0 & \lambda 
\end{array} \right)$, $\lambda>0$. Eq. (\ref{eq:2state}) gives, $\beta (\lambda-b)=0$.
If $\beta=0$, then the agent dies sometime between $t=i$ and $t=k$. If $\lambda=b$ then $S=N$,
the signal from the future could not change the transition probability at $t=i$. The agent
cannot change his own fate by sending a signal to the past.

\subsection{Deutsch's unproven theorem paradox}

An agent sends a signal containing the proof of a mathematical theorem back in
time. The signal is encoded in a document that the agent reads in future time.
Denote the existence of the proof as $\s=0$ and absence of the proof as
$\s=1$. A general form of $V$ is,
$
V = \left( \begin{array}{cc}
\alpha & \beta^* \\
\alpha^* & \beta \end{array} \right),
$
$\alpha^*=1-\alpha$, $\beta^*=1-\beta$.
The basic assumptions we use are (i) the transition from $\s=1$ to $\s=0$ (transition of absence
of proof to existence of proof) happens solely through
the signal traveling back in time, and (ii) the transition from $\s=0$ to $\s=1$ happens
with zero probability (once the proof is obtained, it never gets lost). Hence 
$\beta=1$ and $\alpha^*=0$. The transition probabilities are 
$\hT_i(\s_i=0|\ts_{i-1}=1,\s_k=1) = 0$ representing no signal sent if proof does
not exist at $t=k$ ($\s_k=1$). $\hT_i(\s_i=0|\ts_{i-1}=1,\s_k=0)=1$ represents a signal
being sent when the proof exists at $t=k$. These basic assumptions contradict with 
Eq. (\ref{eq:2state}),
$[\hT_i(0|1,1)-\hT_i(0|1,0)](\alpha^*-\beta) = (0-1)(0-1) \neq 0$.
Hence the assumptions are false and Deutsch's unproven theorem paradox is precluded.

The paradox can be resolved if we relax the assumptions. Suppose we allow the
possibility that the proof can get lost ($\alpha^*\geq 0$) and that the proof can be derived
by some brilliant mathematician $\beta\leq 1$. Then Eq. (\ref{eq:2state}) can be 
satisfied if $\alpha^*=\beta$. There is no paradox here because the proof can be sent
back in time and subsequently be lost. It can be re-derived again and be sent back to
the past.

\begin{figure}
\begin{picture}(180,130)(0,0)
\put(0,-15){\includegraphics[width=6cm]{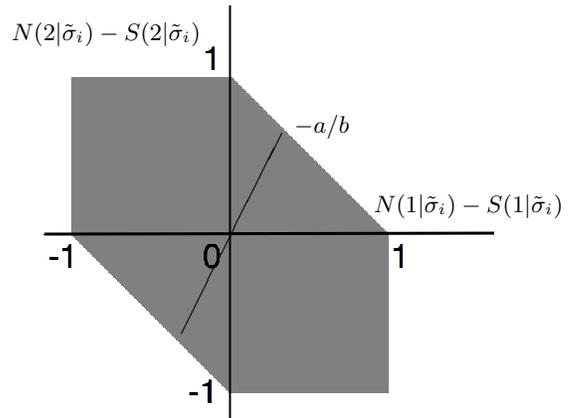}}
\put(125,65){$N(1|\ts_i)-S(1|\ts_i)$}
\put(-12,130){$N(2|\ts_i)-S(2|\ts_i)$}
\put(52.2,19){\line(1,2){38}}
\put(95,95){$-a/b$}
\end{picture}
\caption{Shaded region shows the possible values of 
$N(1|\ts_i,0)-S(1|\ts_i)$, (x-axis) and
$N(2|\ts_i,0)-S(2|\ts_i)$, (y-axis).}
\label{fig:range}
\end{figure}

\subsection{Three-state system}

For a three-state system, $\s$ takes the values $\{0,1,2\}$. For simplicity, let
$\hT_i(\s_i|\ts_{i-1},0)=N(\s_i|\ts_{i-1})$ and 
$\hT_i(\s_i|\ts_{i-1},1)=\hT_i(\s_i|\ts_{i-1},2)=S(\s_i|\ts_{i-1})$. Using 
Eq. (\ref{eq:thatnorm}) and (\ref{eq:TV}), 
\bea 
\label{eq:3state}
[N(1|\ts_{i-1})-S(1|\ts_{i-1})] [V(0|1)-V(0|0)] +\mbox{\hspace{-.1cm}} && \\ \nonumber
[N(2|\ts_{i-1})-S(2|\ts_{i-1})] [V(0|2)-V(0|0)] = && 0
\eea
this is an equation of the form $xa+yb=0$ given $a=[V(0|1)-V(0|0)]$ and 
$b=[V(0|2)-V(0|0)]$, $x=[N(1|\ts_{i-1})-S(1|\ts_{i-1})]$ and 
$y=[N(2|\ts_{i-1})-S(2|\ts_{i-1})]$ can be solved. There are in general infinitely many
solutions. From Eq. (\ref{eq:thatrange}), the range of 
$[N(1|\ts_{i-1})-S(1|\ts_{i-1})]$ and $[N(2|\ts_{i-1})-S(2|\ts_{i-1})]$
is bounded by the shaded region in Fig. \ref{fig:range}.
Given $a$ and $b$, the set of solutions for $x$ and $y$ contains all the points on the line
shown in Fig. \ref{fig:range}. The slope of the line is given by $-a/b$.
$N\neq S$ implies that
transition to the state $\s_i$ depends on future state $\s_k$, that is, signals
from the future can affect the probability distribution of the past.

\subsection{The grandfather paradox in a three-state system}

Consider the three states represent healthy ($\s=2$), sick
($\s=1$) and dead ($\s=0$). First, we lay down our assumptions,
\begin{enumerate}
\item Assume resurrection is impossible so that transition from $\s=0$ to $\s\neq 0$
happens with zero probability. Then the matrix $V$ is of the form,
\be
V = \left( \begin{array}{ccc}
1 & \alpha_0 & \beta_0  \\
0 & \alpha_1 & \beta_1  \\
0 & \alpha_2 & \beta_2 
\end{array} \right)
\ee
with $\alpha_0+\alpha_1+\alpha_2=1$ and $\beta_0+\beta_1+\beta_2=1$. 
\item The agent is able to send a signal back in time to kill himself only if he is
not dead at $t=k$.
\end{enumerate}
$\hT_i(\s_i|\s_{i-1}, 1)$ and $\hT_i(\s_i|\s_{i-1}, 2)$ are the conditional
probabilities that the agent is alive and sends a signal back in time to kill himself.
Let, $\hT_i(\s_i|\ts_{i-1}, 1) = \hT_i(\s_i|\ts_{i-1}, 2) = S(\s_i|\ts_{i-1})$.
$\hT_i(\s_i|\s_{i-1}, 0)$ is the conditional
probability that the agent is dead at $t=k$ and can not send a signal back in time
to kill himself.  Let $\hT_i(\s_i|\ts_{i-1}, 0)=N(\s_i| \ts_{i-1})$.
Hence $S$ (the ``killing" matrix) $N$ are,
\be
S = \left( \begin{array}{ccc}
1 & 1 & 1 \\
0 & 0 & 0 \\
0 & 0 & 0
\end{array} \right)
\mbox{\hspace{.3cm}}
N = \left( \begin{array}{ccc}
1 & a_0 & b_0  \\
0 & a_1 & b_1  \\
0 & a_2 & b_2 
\end{array} \right)
\label{eq:kill3}
\ee
We have from Eq. (\ref{eq:3state}),
\bea
\label{eq:gfp3state}
a_1 (1-\alpha_0) + a_2 (1-\beta_0) & = & 0 \\ \nonumber
b_1 (1-\alpha_0) + b_2 (1-\beta_0) & = & 0 
\eea
There are four cases in which Eq. (\ref{eq:gfp3state}) is satisfied.
\begin{enumerate}
\item $\alpha_0=1$ and $\beta_0=1$. Then $V=S$ which means the agent is dead at $t=k$
with probability 1 (recall that $S$ is the killing matrix).
\item $\alpha_0=1$ and $\beta_0< 1$. To satisfy Eq. (\ref{eq:gfp3state}),
$a_2=b_2=0$. In this case the agent is dead at $t=k$ with 
probability 1 (see Appendix B for the proof).
\item $\alpha_0<1$ and $\beta_0=1$. To satisfy Eq. (\ref{eq:gfp3state}),
$a_1=b_1=0$. In this case the agent is dead at $t=k$ with probability
1 (see Appendix B for the proof).
\item $\alpha_0< 1$ and $\beta_0< 1$. Then $a_1=a_2=b_1=b_2=0$ and $N=S$ which
means the agent is dead at $t=i$ with probability 1.
\end{enumerate}
In all cases, the agent is dead with probability 1 at $t=k$ and hence never has
a chance to send a signal back in time to kill himself. Suppose $S$ is not the killing
matrix (Eq. (\ref{eq:kill3})) or resurrection is possible, then this argument does not
hold, and the agent is able to alter his fate by changing the probability of being
healthy, sick or dead. 

\subsection{The chicken-and-egg paradox}

Consider the chicken-and-egg paradox in which at time $t=k$, a hen travels back in
time to $t=i$ to lay an egg. The egg hatches into the hen herself. At this time point,
there are two copies of the hen, the older self and the younger self (the chick). As
both copies travel to time $t=k$, the chick grow older and travels back in time to
lay the egg.
This paradox seems ``self-consistent" in the sense that there is no contradiction in
existence of the hen and chick from one time point to another. However the problem is
the hen seems to pop out from nowhere.

There are three possible states, hen and chick ($\s=0$), hen only ($\s=1$) and
no hen and no chick ($\s=2$). We exclude the state of chick only, otherwise we would
need four states.

There are no hen and no chick initially, hence $\ts_{i-1}=2$. 
Let $\hT_i(\s_i|\ts_{i-1},1) = \hT_i(\s_i|\ts_{i-1},2)=N(\s_i|\ts_{i-1})$. This is the
case when no chick travels back in time and hence there remains no hen and no chick at
$t=i$. Let 
$\hT_i(\s_i| \ts_{i-1},0) = S(\s_i|\ts_{i-1})$, the chick travels back in time
from $t=k$ to $t=i$. The matrices $S$ and $N$ are,
\be
S = \left( \begin{array}{ccc}
1 & 0 & 0 \\
0 & 1 & 1 \\
0 & 0 & 0 \end{array} \right)
\mbox{\hspace{.4cm}}
N = \left( \begin{array}{ccc}
1 & 0 & 0 \\
0 & 1 & 0 \\
0 & 0 & 1 \end{array} \right)
\ee
The matrix $V$ is of the form,
\be
V = \left( \begin{array}{ccc}
\alpha_0 & \beta_0 & 0 \\
\alpha_1 & \beta_1 & 0 \\
\alpha_2 & \beta_2 & 1 \end{array} \right)
\ee
The first two columns are general expressions with $\sum_{j=0}^2 \alpha_j=1$
and $\sum_{j=0}^2 \beta_j=1$. The last column is $(0,0,1)^T$ because when there is 
no hen and no chick at time $t=i$, then there will be no hen and no chick at $t=k$.
Now consider the probability,
\be
P(\ts_{i-1},\s_i,\s_k) = p(\ts_{i-1})
\hT_i(\s_i|\ts_{i-1},\s_k) V(\s_k|\s_i)
\ee
$p(\ts_{i-1})$ is the probability of sampling the state $\ts_{i-1}$. Since 
$\ts_{i-1}=2$, $p(\ts_{i-1})=\delta_{\ts_{i-1},2}$.
We remind the reader that the probability distribution $V$ is the sum of probabilities
over all possible intermediate sequences. The chicken-and-egg paradox requires both
hen and chick to be present at
$t=k$ ($\s_k=0$) and the chick to appear at $t=i$ ($\s_i=1$), all intermediate states
can take arbitrary values. Reading off entries from matrices $S$ and $V$,
\be
P(\ts_{i-1}=2,\s_i=1,\s_k=0) = \beta_0
\ee
Using Eq. (\ref{eq:3state}) we can calculate what $\beta_0$ should be,
\bea
[ 1-0]  (\beta_0 - \alpha_0) -
[0-1] \alpha_0 & = & 0 \\ \nonumber
\Rightarrow \beta_0 &= & 0
\eea
The sum of probabilities of all possible sequences of states that represent the 
chicken-and-egg paradox equals zero. Therefore the chicken-and-egg event happens with
zero probability.

\section{Discussion}

We have shown, using a graphical model with a loop back into the past, that
the grandfather paradox, Deutsch's unproven theorem paradox and the chicken-and-egg
paradox are precluded in time travel. We have also demonstrated that changing the
probability distributions of the past is possible when no contradicting events
are present. For the paradoxes we discussed in this paper, we gave scenarios in which
the paradoxes are resolved. Our analysis is based on isolated two- and three-state systems.

For future work, it would be useful to generalize our formalism to arbitrary systems.
Lastly, in cases when the causal relationship between events at different times are
very complex, the existence of time travel paradoxes in these cases may be very 
subtle. We hope that our mathematical framework can be used to uncover new time
travel paradoxes, especially those that are embedded in complex interactions of events
and are not obvious.


The author would like to thank Mui Leng Seow and Ivana Mihalek for proofreading this
article.

\section{Appendix A: Derivation of Eq. (\ref{eq:TV})}

Probability of a sequence $\pi_n$ is given by,
\be
P(\pi_n) = p(\s_1)T_2(\s_2|\s_1)\cdots \hT_i(\s_i|\s_{i-1},\s_k)
\cdots T_n(\s_n|\s_{n-1})
\label{eq:mcloop}
\ee
Summing over all sequences,
\bea 
\label{eq:sumseq}
\sum_{\{\pi_n\} } P(& &\pi_n) = \\ \nonumber
\sum_{\s_1,\s_2,\cdots,\s_n} & & p(\s_1) T_2(\s_2|\s_1)\cdots
\hT_i(\s_i|\s_{i-1},\s_k)\cdots T_n(\s_n|\s_{n-1})
\eea
Since $\sum_{\s_j} T_j(\s_j|\s_{j-1})=1 \forall j$, 
summation can be evaluated recursively
between $\s_{k+1}$ and $\s_n$. That is,
\be
\sum_{\s_{k+1},\cdots \s_n} 
T_{k+1}(\s_{k+1}|\s_k) \cdots T_n(\s_n|\s_{n-1}) = 1
\ee
Next define,
\be
U(\s_{i-1}) = \sum_{\s_1,\cdots \s_{i-2}} p(\s_1) T_2(\s_2|\s_1)\cdots
T_{i-1}(\s_{i-1}|\s_{i-2})
\ee
\be
V(\s_k|\s_i) = \sum_{\s_{i+1},\cdots \s_{k-1}} 
T_{i+1}(\s_{i+1}|\s_i)\cdots
T_k(\s_k|\s_{k-1})
\ee
Then Eq. (\ref{eq:sumseq}) becomes,
\be
\sum_{\{ \pi_n\} } P(\pi_n) = 
\sum_{\s_{i-1},\s_i,\s_k} 
U(\s_{i-1}) \hT_i(\s_i | \s_{i-1},\s_k)
V(\s_k|\s_i )
\ee
The objective is to find the conditions in which $\sum_{\{\pi_n\} } P(\pi_n)=1$.
$U(\s_{i-1})$ is the probability of sampling the state $\s_{i-1}$, it depends
on the conditional probabilities $T_j$, $j\leq i-1$ and the initial condition
$p(\s_1)$.
We therefore have the 
freedom to choose $U$ for example, by choosing different initial
conditions. Holding $T$ and $V$ fixed, we require $\sum P(\pi_n)=1$ for different
choices of $U$, in which we arrive at,
\be
\sum_{\s_i,\s_k} 
\hT_i(\s_i | \ts_{i-1},\s_k)
V(\s_k|\s_i ) = 1
\ee

\section{Appendix B: The grandfather paradox in a three-state system}

We present the proof that for the grandfather paradox in a three-state system, 
the probability that the agent is dead at $t=k$ is one. We consider cases II and III
in which Eq. (\ref{eq:gfp3state}) is satisfied,

\subsection{Case II: $\alpha_0=1$ and $\beta_0<1$}

In this case, $a_2=b_2=0$ and,
\be
V=\left( \begin{array}{ccc}
1 & 1 & \beta_0 \\
0 & 0 & \beta_1 \\
0 & 0 & \beta_2 
\end{array} \right)
\label{eq:B:V1}
\ee
\be
N=\left( \begin{array}{ccc}
1 & a_0 & b_0 \\
0 & a_1 & b_1 \\
0 & 0 & 0
\end{array} \right)
\label{eq:B:N1}
\ee
We calculate the probability that the agent is dead,
\bea \nonumber
P(\s_k=0) & = & \sum_{\s_1,\cdots,\s_{k-1}} p(\s_1) T_2(\s_2|\s_1) \cdots \\ \nonumber
& = & \sum_{\s_{i-1},\s_i} U(\s_{i-1}) \hT_i(\s_i|\s_{i-1},0) V(0|\s_i) \\
& = & \sum_{\s_{i-1},\s_i} U(\s_{i-1}) N(\s_i|\s_{i-1}) V(0|\s_i) 
\label{eq:B:p}
\eea
Reading off the entries of matrices $V$ and $N$ in Eq. (\ref{eq:B:V1}) and (\ref{eq:B:N1}),
we get $\sum_{\s_i} N(\s_i|\s_{i-1}) V(0|\s_i) = 1$ for all $\s_{i-1}$. Hence $P(\s_k=0)=1$.

\subsection{Case III: $\alpha_0<1$ and $\beta_0=1$}
In this case, $a_1=b_1=0$ and,
\be
V=\left( \begin{array}{ccc}
1  & \alpha_0 & 1\\
0  & \alpha_1 & 0\\
0  & \alpha_2 & 0
\end{array} \right)
\label{eq:B:V2}
\ee
\be
N=\left( \begin{array}{ccc}
1 & a_0 & b_0 \\
0 & 0 & 0 \\
0 & a_2 & b_2 \\
\end{array} \right)
\label{eq:B:N2}
\ee
We calculate the probability that the agent is dead, using Eq. (\ref{eq:B:p})
and reading off the entries of matrices $V$ and $N$ in Eq. (\ref{eq:B:V2}) and (\ref{eq:B:N2}),
we get $\sum_{\s_i} N(\s_i|\s_{i-1}) V(0|\s_i) = 1$ for all $\s_{i-1}$. Hence $P(\s_k=0)=1$.

\end{document}